\documentclass[12pt,draftclc,onecolumn]{IEEEtran}
\usepackage{siunitx}
\ifCLASSINFOpdf
   \usepackage[pdftex]{graphicx}
 \else
\fi
\usepackage{amssymb, latexsym}
\usepackage{amsmath} 
\usepackage{amsthm}
\usepackage{verbatim}   % useful for program listings
\usepackage{color}      % use if color is used in text
\usepackage{mathtools}
\usepackage[dvipsnames]{xcolor}
\usepackage{color,soul}

\definecolor{mygray}{gray}{0.5}

\begin{document}
\title{On the retrieval of forward-scattered waveforms from acoustic reflection and transmission data with the Marchenko equation}

\author{Joost~van~der~Neut,~Joeri~Brackenhoff,~Giovanni~Meles,\
Lele~Zhang,~Evert~Slob,~Kees~Wapenaar}

\maketitle

\begin{center}{\today}\end{center}

\hfill \break

\noindent J. van der Neut, E. Slob and K. Wapenaar are with the Department of Geoscience and Engineering, Delft University of Technology, 2600 GA, Delft, The Netherlands. \\
\noindent J. Brackenhoff is with ETH, 8092, Z\"{u}rich, Switzerland. \\
\noindent G. Meles is with University of Lausanne, 1015, Lausanne, Switzerland. \\
\noindent L. Zhang is with China University of Geosciences, No. 388 Lumo Road, Wuhan, China.

\newpage

\begin{abstract}
A Green's function in an acoustic medium can be retrieved from reflection data by solving a multidimensional Marchenko equation. This procedure requires \textit{a-priori} knowledge of the initial focusing function, which can be interpreted as the inverse of a transmitted wavefield as it would propagate through the medium, excluding (multiply) reflected waveforms. In practice, the initial focusing function is often replaced by a time-reversed direct wave, which is computed with help of a macro velocity model. Green's functions that are retrieved under this (direct-wave) approximation typically lack forward-scattered waveforms and their associated multiple reflections. We examine whether this problem can be mitigated by incorporating transmission data. Based on these transmission data, we derive an auxiliary equation for the forward-scattered components of the initial focusing function. We demonstrate that this equation can be solved in an acoustic medium with mass density contrast and constant propagation velocity. By solving the auxiliary and Marchenko equation successively, we can include forward-scattered waveforms in our Green's function estimates, as we demonstrate with a numerical example.

\end{abstract}

%%%%%%%%%%%%%%%%%%%%%%%%%%%%%%%%%%%%%%%%%%%%%%%%%%%%%%%%%%%%%%%%%
\section{Introduction}

It has been shown that the Green's function between a horizontal acquisition surface and an arbitrary location ${\bf{x}}$  inside an unknown lossless acoustic medium can be retrieved from a single-sided reflection response by solving a multidimensional Marchenko equation \cite{wapenaar13}. This insight has led to numerous applications in the field of applied geophysics; see \cite{wapenaar21a} for an overview. Besides knowledge of the single-sided reflection response at an acquisition surface, the Marchenko methodology requires access to the source signature, which can sometimes be retrieved from the recorded data \cite{mildner19a}, and an initial estimate of the transmitted wavefield as it would propagate from the acquisition surface to ${\bf{x}}$  in absence of (multiple) reflections. Typically, this initial estimate is obtained from a macro model of the propagation velocity \cite{broggini14}. The phase \cite{mildner17,chen20} and amplitude \cite{mildner19b} of the initial estimate can be updated within the Marchenko framework. 

In theory, the wavefield that is used to initialize the Marchenko scheme should include all forward-scattered waveforms \cite{vasconcelos15,diekmann21}, which are those waveforms that do not (ever) change vertical direction, while propagating from the (horizontal) acquisition boundary to ${\bf{x}}$. In practice, we typically compute the initial focusing function in a smooth macro model, which does not contain sharp contrasts. Consequently, forward-scattered waveforms and their associated multiples will not be accurately reconstructed \cite{neut15}, which can harm the (Marchenko) imaging process \cite{vargas21}. Another problem is posed by thin-layered structures, generating multiple reflections with short periods that cannot be resolved due to the finite frequency content of the data \cite{slob14}. This problem can be mitigated (at least to some extent) by enforcing energy conservation and minimum-phase conditions \cite{dukalski19,elison20,peng21a}.

The Marchenko scheme can also be applied to ultrasonic data \cite{wapenaar18,cui18,costa18a}, opening new ways for biomedical applications, especially below objects with strong contrast such as the human skull \cite{meles19}. Hence, the Marchenko equation might be tailored to supplement common biomedical modalities, such as transcranial wavefield focusing \cite{clement02}, brain imaging \cite{guasch20} or transcranial photoacoustics \cite{poudel20a}. Remarkably, some of the acquisition designs that are common for these applications allow the collection of auxiliary transmission data. As the transmission response bears an imprint of the desired forward-scattered waveforms, these data could be key to improve the initial wavefield estimate which is needed to solve the Marchenko equation. In this paper, we elaborate on this idea. Unlike the Marchenko scheme of \cite{kiraz21}, which has been proposed recently for closed-boundary data, we make a sharp distinction between reflection and transmission data. We start with a brief derivation of the Marchenko equation for reflection data. By modifying the derivation slightly, we find an auxiliary equation for the recorded transmissions. Our aim is to resolve forward-scattered waveforms from this equation, in order to improve the initial wavefield that is used in the Marchenko equation. In this paper, we consider a medium with density contrast and constant propagation velocity. By solving the auxiliary (transmission-based) equation and the (reflection-based) Marchenko equations successively, forward-scattered waveforms and their associated multiples can be included in our Green's function estimates, as we demonstrate numerically. As an example of a potential application for our methodology, we consider Marchenko-based solutions of inverse source problems, which are key in photoacoustic imaging \cite{huang13}. Applications in media with significant velocity contrast are more challenging and require additional research to be conducted.

\begin{figure}
\centering
{\scalebox{0.60}[0.60]{\includegraphics[trim={0 0 0 0},clip]{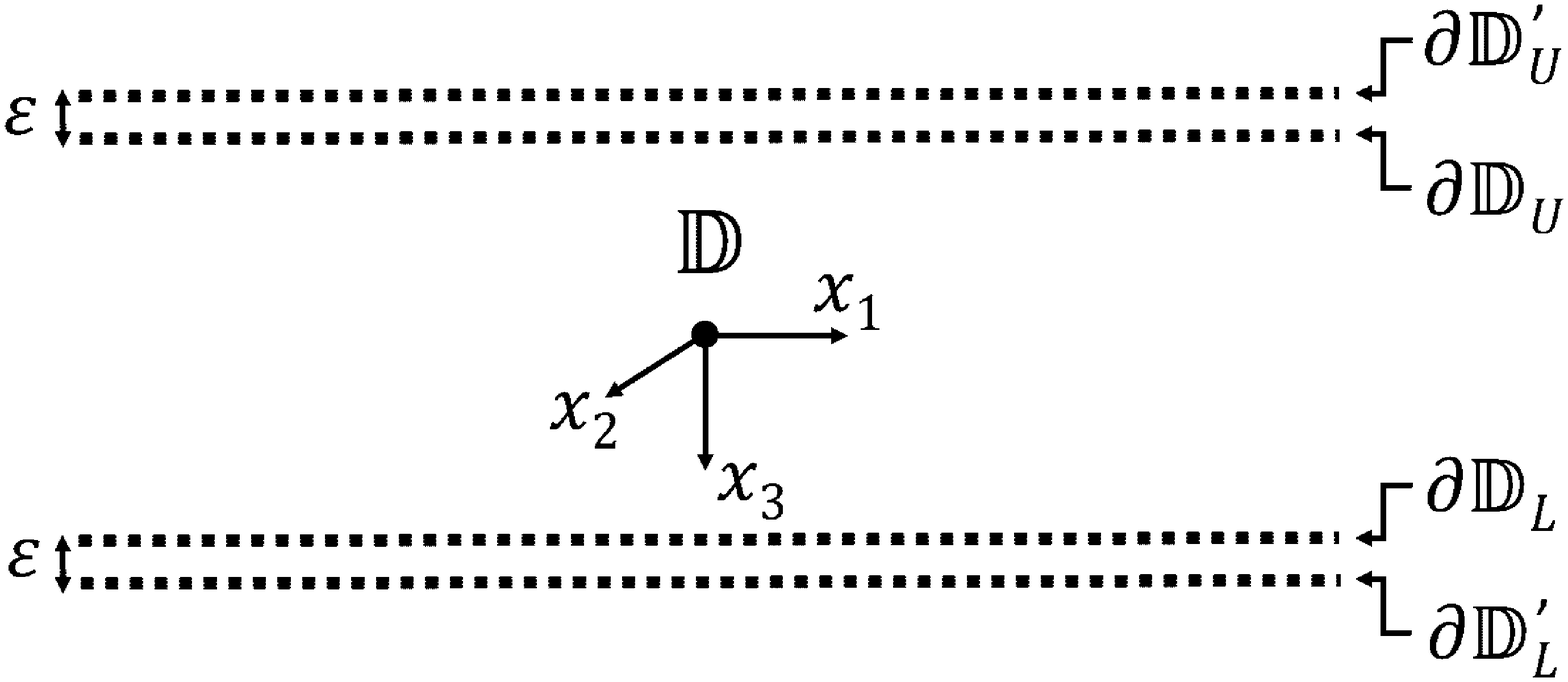}}}
\caption{Configuration: Volume $\mathbb{D}$ is enclosed by horizontal boundaries $\partial \mathbb{D}_U$ and $\partial \mathbb{D}_L$ (both extending infinitely in the lateral directions). The coordinate system is also indicated. The medium is non-reflective above $\partial \mathbb{D}_U$.  Vertical dipole sources are located at an upper boundary $\partial \mathbb{D}_U^\prime$, at an infinitesimal distance $\epsilon \rightarrow 0$ above $\partial \mathbb{D}_U$, and at the lower boundary $\partial \mathbb{D}_L^\prime$, at an infinitesimal distance $\epsilon \rightarrow 0$ below $\partial \mathbb{D}_L$. Receivers are located at $\partial \mathbb{D}_U$.
\label{fig:n1}}
\end{figure}

\section{Marchenko equation for reflection data}
\label{reflection}

In this section, we briefly review a recent derivation of the Marchenko-type representation by \cite{wapenaar21} for Green's function retrieval from reflection data. Consider the configuration in Fig. \ref{fig:n1}. Let ${\bf{x}}=\left( x_1, x_2, x_3 \right)$ be a location in 3D space, where the $x_3$-axis is pointing downwards. Volume $\mathbb{D}$ is bounded by the horizontal boundaries $\partial \mathbb{D}_U$ and $\partial \mathbb{D}_L$, which are located at depth levels $x_{3,U}$ and $x_{3,L}$, respectively.  A lossless acoustic medium is characterized by the propagation velocity $c \left({\bf{x}} \right)$ and the mass density $\rho \left ({\bf{x}} \right)$. We emphasize that all of our methodologies could be extended to include a free surface at the upper boundary \cite{singh17}, or more generally, to allow arbitrary medium properties above this level \cite{ravasi17}. An acoustic pressure field $p\left({\bf{x}},t\right)$ can be expressed as a function of space ${\bf{x}}$ and time $t$. This field can be transformed to the frequency domain by the Fourier transform

\begin{equation}
p \left( {\bf{x}} , \omega \right)
=
\int_{- \infty}^{+ \infty}
p \left( {\bf{x}} , t \right)
e^{i \omega t } 
{\rm d} t
,
\label{eq:S1}
\end{equation}
where $\omega$ is the angular frequency. Wave propagation is assumed to obey the acoustic wave equation

\begin{equation}
\mathcal{L}
p
=
i \omega
q
,
\label{eq:S2}
\end{equation}
with $q \left( {\bf{x}} , \omega \right)$ being a volume-injection rate density source. Further, operator $\mathcal{L}$ is defined as

\begin{equation}
\mathcal{L}
=
\partial_i
\frac{1}{\rho}
\partial_i
+
\frac{\omega^2}{\rho c^2}
,
\label{eq:S3}
\end{equation}
where $\partial_i$ is the spatial derivative in the $i$-direction and Einstein's summation convention applies. Let volume $\mathbb{D}$ be source-free, such that $\forall {\bf{x}} \in \mathbb{D} : q \left( {\bf{x}}, \omega \right) = 0$. We assume that the wavefield is recorded at $\partial \mathbb{D}_U$, where it can be decomposed into downgoing constituents $p^+$ and upgoing constituents $p^-$, such that $p=p^++p^-$. It has been shown that the wavefield at any location ${\bf{x}} \in \mathbb{D}$  may then be expressed as \cite{wapenaar21}

\begin{equation}
\begin{split}
p \left( {\bf{x}} , \omega \right)
 =
\int_{\partial \mathbb{D}_U}
F_U \left( {\bf{x}},{\bf{x}}_U, \omega \right)
p^- \left( {\bf{x}}_U, \omega \right)
{\rm d} {\bf{x}}_U
+
\int_{\partial \mathbb{D}_U}
F_U^\star \left( {\bf{x}},{\bf{x}}_U, \omega \right)
p^+ \left( {\bf{x}}_U, \omega \right)
{\rm d} {\bf{x}}_U
.\end{split}
\label{eq:S4}
\end{equation}
In this representation, superscript $\star$ denotes complex conjugation and it is assumed that evanescent waves at and above $\partial \mathbb{D}_U$ can be neglected. Further, $F_U$ is a so-called focusing function, which focuses at the upper boundary and obeys wave equation (\ref{eq:S2}) with $q=0$. This function is subject to the focusing condition \cite{wapenaar21}

\begin{equation}
\left.
F_U \left( {\bf{x}}, {\bf{x}}_U , \omega \right)
\right|_{x_3 = x_{3,U}}
=
\delta
\left( {\bf{x}}_H - {\bf{x}}_{H,U} \right)
,
\label{eq:S5}
\end{equation}
where $F_U$ is upgoing at and above $\partial \mathbb{D}_U$. In (\ref{eq:S5}), $\delta$ is a (two-dimensional) Dirac delta distribution and ${\bf{x}}_H = \left( x_1, x_2 \right)$ denotes the horizontal coordinates. For the wavefield $p$, we assume that a vertical dipole source is located at ${\bf{x}}_U^\prime \in \partial \mathbb{D}_U^\prime$, just above $\partial \mathbb{D}_U$ (see Fig. \ref{fig:n1}). This results in the (dipole) Green's function

\begin{equation}
\Gamma \left( {\bf{x}}, {\bf{x}}_U^\prime , \omega \right)
=
\frac{-2}{i \omega \rho_U^\prime}
\partial_{3,U}^\prime
G \left( {\bf{x}}, {\bf{x}}_U^\prime, \omega \right)
\label{eq:S6}
,
\end{equation}
where $\partial_{3,U}^\prime$ is a vertical partial derivative applied at ${\bf{x}}_U^\prime$ and $\rho_U^\prime$ is the density at $\partial \mathbb{D}_U^\prime$. In (\ref{eq:S6}), $G\left( {\bf{x}}, {\bf{x}}_U^\prime, \omega \right)$ is a Green's function of a monopole source at ${\bf{x}}_U^\prime$, evaluated at ${\bf{x}}$, obeying (\ref{eq:S2}) with $q = \delta \left( {\bf{x}} - {\bf{x}}_U^\prime \right)$. When ${\bf{x}}_U^\prime$ approaches $\partial \mathbb{D}_U$ in the limit $\epsilon = x_{3,U} - x_{3,U}^\prime \rightarrow 0$ (see Fig. \ref{fig:n1}), it can be deduced that the downgoing part of the dipole response obeys \cite{wapenaar17}

\begin{equation}
\left.
\lim_{x^\prime_{3,U} \rightarrow {x_{3,U}}}
\Gamma^+ \left( {\bf{x}}, {\bf{x}}_U^\prime , \omega \right)
\right|_{x_3=x_{3,U}}
=
\delta \left( {\bf{x}}_H - {\bf{x}}_{H,U}^\prime \right)
\label{eq:S7}
.
\end{equation}
When we substitute $p \left( {\bf{x}}, \omega \right) = \Gamma \left( {\bf{x}}, {\bf{x}}_U^\prime, \omega \right)$ into (\ref{eq:S4}) and apply (\ref{eq:S7}), it follows that

\begin{equation}
\begin{split}
\Gamma \left( {\bf{x}} , {\bf{x}}_U^\prime, \omega \right)
=
\int_{\partial \mathbb{D}_U}
F_U \left( {\bf{x}},{\bf{x}}_U, \omega \right)
\Gamma^- \left( {\bf{x}}_U, {\bf{x}}_U^\prime, \omega \right)
{\rm d} {\bf{x}}_U
+
F_U^\star \left( {\bf{x}},{\bf{x}}_U^\prime, \omega \right)
.\end{split}
\label{eq:S8}
\end{equation}
In this representation, $\Gamma^- \left( {\bf{x}}_U, {\bf{x}}_U^\prime, \omega \right) $ can be interpreted as the (upgoing) reflection response of the medium recorded at ${\bf{x}}_U$, stemming from a dipole source at ${\bf{x}}_U^\prime$. We wish to express this result in the time domain with help of the inverse Fourier transform, which is defined for an arbitrary wavefield as the inverse of (\ref{eq:S1}); that is \cite{bracewell00}

\begin{equation}
p \left( {\bf{x}}, t \right)
=
\frac{1}{\pi}
\Re
\left[
\int_0^\infty
p \left( {\bf{x}}, \omega \right)
e^{- i \omega t}
{\rm d} \omega
\right]
,
\label{eq:S9}
\end{equation}
where it is assumed that $p \left( {\bf{x}}, t \right)$ is real-valued and $\Re$ denotes the real part. With help of these definitions, (\ref{eq:S8}) can be rewritten in the time domain as

\begin{equation}
\Gamma_U
=
\left(
\mathcal{R}_{U}+ \mathcal{Z}
\right)
F_U
.
\label{eq:S10}
\end{equation}
In this expression, we have $\Gamma_U= \Gamma \left( {\bf{x}} , {\bf{x}}_U^\prime, t \right)$, while $\mathcal{R}_{U}$ is an operator for multidimensional convolution with the reflection response $\Gamma^- \left( {\bf{x}}_U, {\bf{x}}_U^\prime, t \right)$ at the upper boundary, obeying

\begin{equation}
\mathcal{R}_{U}
 F_U
=
\int_{\partial \mathbb{D}_U}
F_U \left( {\bf{x}},{\bf{x}}_U, t \right)
\ast
\Gamma^- \left( {\bf{x}}_U, {\bf{x}}_U^\prime, t \right)
{\rm d} {\bf{x}}_U
,
\label{eq:S11}
\end{equation}
where $\ast$ denotes temporal convolution. Further, $\mathcal{Z}$ is an operator for time reversal.

Our goal is to retrieve the focusing function from (\ref{eq:S10}). To achieve this goal, we require some prior knowledge about the Green's function $\Gamma \left( {\bf{x}}, {\bf{x}}_U^\prime,t \right)$. More specifically, we assume that, for each $\left( {\bf{x}}, {\bf{x}}_U^\prime \right)$-pair, the traveltime $t_{Ud} \left( {\bf{x}}, {\bf{x}}_U^\prime \right)$ of the first (or direct) arrival of this Green's function can be estimated  from a macro velocity model \cite{broggini14} (in case of a triplicated wave, $t_{Ud}$ is the traveltime of the first onset \cite{wapenaar14}). Based on these traveltimes, we design a window operator $\Theta_U$ (also referred to as a projector \cite{dukalski17}) that removes all arrivals at  $t \geq t_{Ud} - t_\epsilon$ (note that our window is not symmetric in time, in contrast to various previous publications). Here, subscript $U$ refers to the upper boundary, where the window operator is applied. In our formulation, a small additional time-shift  $t_\epsilon$  has been included to account for the finite frequency content of the data. In practice, we typically choose $t_\epsilon$ as half the temporal support of the source wavelet \cite{thorbecke17}. Based on causality, we assume that

\begin{equation}
\Theta_U
\Gamma_U
=
0
.
\label{eq:S12}
\end{equation}
In various publications on geophysical applications of the Marchenko equation, the medium is assumed to be layered with moderately curved interfaces \cite{wapenaar13,wapenaar21a}. Under these conditions, the focusing function consists of a time-reversed direct wave, which is timed at $-t_{Ud}$, and a coda, which is timed thereafter. A common interpretation is that the direct wave focuses at ${\bf{x}}$ when injected into the medium from the upper boundary \cite{wapenaar17}, while the coda is associated with all (primary and multiple) reflections that are generated  between this boundary and ${\bf{x}}$. In media with increasing propagation velocity (which are common in geophysical settings), problems arise at long offsets, due to incorrect handling of refracted waves and post-critical reflections \cite{diekmann21,zhang19}. In the presence of sharp discontinuities in the lateral direction, such as point diffractors, the focusing function contains additional forward-scattered components (i.e. waveforms that have not altered their vertical propagation direction between the upper boundary and ${\bf{x}}$) that are (partly) timed before $-t_{Ud}$ \cite{vasconcelos15,neut15,vargas21}. To allow these (unknown) components in our formulation, we formally partition the focusing function in an initial focusing function $F_{Ui}$, containing all waveforms in the interval $\left(-\infty,-t_{Ud}+t_\epsilon \right]$, and a coda $F_{Um}$, containing all waveforms in the interval $\left(-t_{Ud}+t_\epsilon,\infty \right)$. With help of these definitions, we may write

\begin{equation}
F_U
=
F_{Ui} + F_{Um}
.
\label{eq:S13}
\end{equation}
When the operators $\mathcal{Z}$ and $\Theta_U$ are applied successively to the focusing function, it follows from these definitions that

\begin{equation}
\Theta_U \mathcal{Z}  F_U
=
\Theta_U \mathcal{Z} F_{Um}
.
\label{eq:S14}
\end{equation}
When we substitute (\ref{eq:S13}) into (\ref{eq:S10}) and apply operator $\Theta_U$ (from the left) to both sides of the result, it follows with help of (\ref{eq:S12}) and (\ref{eq:S14}) that

\begin{equation}
-
\Theta_U 
\mathcal{R}_{U} 
F_{Ui}
=
\Theta_U \left( \mathcal{R}_{U} + \mathcal{Z} \right)
 F_{Um} 
 .
\label{eq:S15}
\end{equation}
This result is generally known as the Marchenko equation. If the initial focusing function $F_{Ui}$ is known, (\ref{eq:S15}) can be solved for the coda $F_{Um}$. It is common practice to approximate the initial focusing function $F_{Ui}$  for any relevant $\left( {\bf{x}},{\bf{x}}_U^\prime \right)$-pair by a time-reversed direct wave $F_{Ud}$, which is typically computed in an approximate macro velocity model \cite{broggini14}.  We refer to this practice as the direct-wave approximation. Under this approximation, we ignore additional waveforms $F_{Ua}$ that are (mostly) related to forward-scattering, which we define formally as

\begin{equation}
F_{Ua} = F_{Ui} - F_{Ud}.
\label{eq:S16}
\end{equation}
Substitution of $F_{Ui} = F_{Ud} + F_{Ua}$ into (\ref{eq:S15}) yields

\begin{equation}
-
\Theta_U 
\mathcal{R}_{U} 
\left( F_{Ud} + F_{Ua} \right)
=
\Theta_U \left( \mathcal{R}_{U} + \mathcal{Z} \right)
 F_{Um} 
.
\label{eq:S17}
\end{equation}

By assuming $F_{Ua}=0$ (i.e. the direct-wave approximation), the coda $F_{Um}$ can be resolved from (\ref{eq:S17}) by linear inversion. A common strategy for the inversion is to rewrite the equation by a Neumann series expansion, which is guaranteed (at least in 1D for infinite frequency content) to converge as long as the spectral radius of operator $\mathcal{R}_{U}$ is less than one \cite{dukalski17,slob21}. However, a variety of alternative numerical solvers might be employed \cite{dukalski17,slob21,santos21,ravasi21}. For the construction of operator $\mathcal{R}_{U}$, we require access to a complete, well-sampled reflection response \cite{jia18,peng21} and sufficient aperture \cite{sripanich19b}. Modifications of the methodology have been proposed to allow for gaps in the acquisition design \cite{haindl21} and imperfect sampling \cite{ijsseldijk21}. Once the coda $F_{Um}$ is resolved, the complete focusing function can be constructed with (\ref{eq:S13}), and eventually the Green's function follows from (\ref{eq:S10}). In this section, we have derived a Marchenko equation for (vertical) dipole Green's functions, see (\ref{eq:S6}). However, the theory can be modified for the retrieval of monopole Green's functions, see \cite{wapenaar21}.

\section{Auxiliary equation for transmission data}
\label{transmission}

It is well-known that forward-scattered waveforms can not be accurately retrieved under the direct-wave approximation \cite{vasconcelos15,diekmann21,neut15,vargas21}. Ideally, we would like to add the additional components $F_{Ua}$ to $F_{Ud}$, prior to solving the Marchenko equation (\ref{eq:S17}). We show in the following that, when auxiliary transmission data ara available, some of these components can be recovered. We acquire these data by placing additional dipole sources at the lower boundary $\partial \mathbb{D}_L^\prime$ (which is located just below $\partial \mathbb{D}_L$, see Fig. \ref{fig:n1}). We define their associated (dipole) Green's functions as

\begin{equation}
\Gamma \left( {\bf{x}}, {\bf{x}}_L^\prime , \omega \right)
=
\frac{-2}{i \omega \rho_L^\prime}
\partial_{3,L}^\prime
G \left( {\bf{x}}, {\bf{x}}_L^\prime, \omega \right)
\label{eq:S18}
,
\end{equation}
with ${\bf{x}}_L^\prime \in \partial \mathbb{D}_L^\prime$, $\partial_{3,L}^\prime$ denoting the vertical partial derivative at ${\bf{x}}_L^\prime$ and $\rho_L^\prime$ being the density at $\partial \mathbb{D}_L^\prime$. Here, $G \left( {\bf{x}}, {\bf{x}}_L^\prime, \omega \right)$ is a monopole Green's function, obeying (\ref{eq:S2}) with $q = \delta \left( {\bf{x}}-{\bf{x}}_L^\prime \right)$. When we substitute $p = \Gamma \left( {\bf{x}},{\bf{x}}_L^\prime, \omega \right)$   into (\ref{eq:S4}), it follows $\forall {\bf{x}} \in {\mathbb{D}}$ that

\begin{equation}
\Gamma \left( {\bf{x}}, {\bf{x}}_L^\prime , \omega \right)
=
\int_{\partial \mathbb{D}_U}
F_U \left( {\bf{x}}, {\bf{x}}_U , \omega \right)
\Gamma^- \left( {\bf{x}}_U, {\bf{x}}_L^\prime , \omega \right)
{\rm d} {\bf{x}}_U
\label{eq:S19}
,
\end{equation}
where we used the fact that the medium is non-reflective above $\partial \mathbb{D}_U$, such that $\left. \Gamma^+ \left( {\bf{x}},{\bf{x}}_L^\prime, \omega \right) \right|_{x_3 = x_{3,U}}=0$. In this representation, $\Gamma^- \left( {\bf{x}}_U, {\bf{x}}_L^\prime, \omega \right)$ can be interpreted as the (upgoing) transmission response of the medium recorded at ${\bf{x}}_U$, stemming from a dipole source at ${\bf{x}}_L^\prime$. After inverse Fourier transformation, (\ref{eq:S19}) can be compactly rewritten as

\begin{equation}
\Gamma_L
=
\mathcal{T}_{LU} F_U
,
\label{eq:S20}
\end{equation}
where $\Gamma_L = \Gamma\left( {\bf{x}}, {\bf{x}}_L^\prime , t \right)$. Further, $\mathcal{T}_{LU}$ is an operator for multidimensional convolution with the transmission response, obeying

\begin{equation}
\mathcal{T}_{LU}  F_U
=
 \int_{\partial \mathbb{D}_U}
F_U \left( {\bf{x}}, {\bf{x}}_U , t \right)
\ast
\Gamma^- \left( {\bf{x}}_U, {\bf{x}}_L^\prime , t \right)
{\rm d} {\bf{x}}_U.
\label{eq:S21}
\end{equation}
Once more, we assume that the traveltimes $t_{Ld} \left( {\bf{x}}, {\bf{x}}_L^\prime \right)$ of the first (or direct) arrivals of $\Gamma \left( {\bf{x}}, {\bf{x}}_L^\prime, t \right)$ can be estimated from a macro velocity model, such that an operator $\Theta_L$ can be constructed which mutes all arrivals at $t \geq t_{Ld} - t_\epsilon$ (where $t_\epsilon$ is a small timeshift, as defined earlier and subscript $L$ denotes the lower boundary, where the window operator is applied). Akin to (\ref{eq:S12}), causality leads to the assumption that

\begin{equation}
\Theta_L
\Gamma_L
=
0
.
\label{eq:S22}
\end{equation}
When we apply operator $\Theta_L$ to (\ref{eq:S20}), it follows straight from (\ref{eq:S22}) that $F_U$ should be in the nullspace of operator $\Theta_L \mathcal{T}_{LU}$:

\begin{equation}
\Theta_L
\mathcal{T}_{LU}
F_U
=0.
\label{eq:S23}
\end{equation}
We may substitute $F_U = F_{Ud} + F_{Ua} + F_{Um}$ into (\ref{eq:S23}) and rewrite the result strategically as

\begin{equation}
-
\Theta_L
\mathcal{T}_{LU}
\left( F_{Ud} + F_{Um} \right)
=
\Theta_L
\mathcal{T}_{LU}
F_{Ua}
.
\label{eq:S24}
\end{equation}
We refer to (\ref{eq:S24}) as our auxiliary equation for transmission data. In this paper, we investigate if this equation can be solved in a medium with density contrast and constant propagation velocity. At first glance, the left-hand side of (\ref{eq:S24}) seems to depend on both $F_{Ud}$ and $F_{Um}$. However, since the waveforms in $F_{Um}$ are timed after the time-reversed direct wave, they reside mainly in the nullspace of $\Theta_L \mathcal{T}_{LU}$. Therefore, we assume that $-\Theta_L \mathcal{T}_{LU} F_{Um}\approx 0$. Consequently, (\ref{eq:S24}) can be rewritten / approximated as

\begin{equation}
-
\Theta_L
\mathcal{T}_{LU}
 F_{Ud}
=
\Theta_L
\mathcal{T}_{LU}
F_{Ua}
.
\label{eq:S25}
\end{equation}

\begin{figure}
\centering
{\scalebox{0.60}[0.60]{\includegraphics[trim={0 0 0 0},clip]{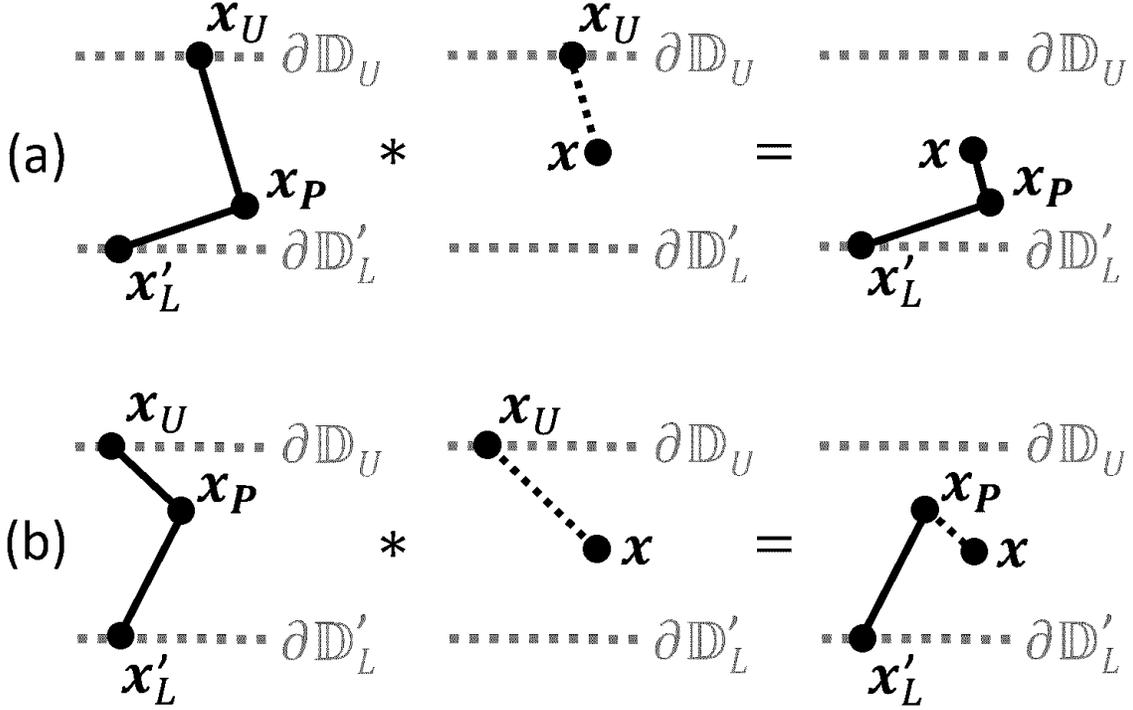}}}
\caption{Let ${\bf{x}}_P$ be a scattering point in a medium with constant propagation velocity, generating a forward-scattered event with traveltime $ t_d \left( {\bf{x}}_U,{\bf{x}}_P \right) + t_d \left( {\bf{x}}_P, {\bf{x}}_L^\prime \right)$ (where ${\bf{x}}_L^\prime$ is a source at $\partial \mathbb{D}_L^\prime$ and ${\bf{x}}_U$ is a receiver at $\partial \mathbb{D}_U$). (a) Situation where $x_{3,P} > x_3$. When the transmission data are convolved with the direct focusing function $F_{Ud} \left( {\bf{x}}, {\bf{x}}_U ,  t \right)$, a physical event is generated with traveltime $t_d \left( {\bf{x}}_P, {\bf{x}}_L^\prime \right) + t_d \left( {\bf{x}}_P, {\bf{x}} \right)$. This process is illustrated in the figure, where the solid and dashed black lines indicate causal and acausal raypaths, respectively (the dashed grey lines designate $\partial \mathbb{D}_U$ and $\partial \mathbb{D}^\prime_L$, while the black dots indicate specific locations).  From the triangle inequality, it follows that $t_d \left( {\bf{x}}_P,{\bf{x}}_L^\prime \right) + t_d \left( {\bf{x}}_P, {\bf{x}} \right) \geq t_d \left( {\bf{x}}, {\bf{x}}_L^\prime \right)$. Hence, the event maps at or after the direct arrival of $\Gamma_L = \Gamma \left( {\bf{x}}, {\bf{x}}_L^\prime,t \right)$, resulting in $\Theta_L \mathcal{T}_{LU} F_{Ud} = 0$. (b)  Situation where $x_{3,P} < x_3$. Now, the generated event is a non-physical arrival with traveltime $t_d \left( {\bf{x}}_P, {\bf{x}}_L^\prime \right) - t_d \left( {\bf{x}}_P, {\bf{x}} \right)$. From the triangle inequality, it follows that $t_d \left( {\bf{x}}_P, {\bf{x}}_L^\prime \right) \leq t_d \left( {\bf{x}}_P, {\bf{x}} \right) + t_d \left( {\bf{x}}, {\bf{x}}_L^\prime \right)$, or $t_d \left( {\bf{x}}_P,{\bf{x}}_L^\prime \right) - t_d \left( {\bf{x}}_P, {\bf{x}} \right) \leq t_d \left( {\bf{x}}, {\bf{x}}_L^\prime \right)$. Hence, the event maps before or at the direct arrival of $\Gamma_L = \Gamma \left( {\bf{x}}, {\bf{x}}_L^\prime,t \right)$, resulting in $\Theta_L
\mathcal{T}_{LU}
F_{Ud}
\neq
0$ as long as ${\bf{x}}_P$, ${\bf{x}}$ and ${\bf{x}}_L^\prime$ are not collinear. In the collinear case, $F_{Ua}$ overlaps with $F_{Ud}$ (i.e. the propagation direction is not altered by scattering at ${\bf{x}}_P$) and cannot be recovered.
\label{fig:n2}}
\end{figure}

In Fig. \ref{fig:n2}(a), we show that the presence of forward-scattered waveforms which are generated below ${\bf{x}}$ yield $-\Theta_L \mathcal{T}_{LU} F_{Ud} = 0$. On the other hand, when forward-scattered waveforms are generated above ${\bf{x}}$, we find $-\Theta_L \mathcal{T}_{LU} F_{Ud} \neq 0$, as illustrated in Fig. \ref{fig:n2}(b). Based on (\ref{eq:S25}), these data should match $\Theta_L \mathcal{T}_{LU} F_{Ua}$, resulting in a linear inverse problem that can be solved for $F_{Ua}$, for instance by LSQR \cite{paige82}. Since $F_{Ui}$ and $F_{Ud}$ are only allowed to be non-zero at $\left( -\infty, -t_{Ud}+t_\epsilon \right]$, so does $F_{Ua}=F_{Ui}-F_{Ud}$, which we enforce during the inversion by restricting the unknown quantity to this time interval. Forward-scattered components can only be resolved if they are kinematically separated from the direct wave, such that they reside outside the nullspace of $\Theta_L \mathcal{T}_{LU}$. The dependence of operator $\Theta_L$ on $t_\epsilon$ reveals that the separation of waveforms that can be recoverd from waveforms that cannot be recovered is intimately related to the frequency content of the data. This observation also means that the transmission loss that forward-scattering imposes on the direct wave should be formally included in our definition of $F_{Ud}$. In practice, we neglect these effects by approximating $F_{Ud}$ in a macro velocity model, which could lead to amplitude mismatches in the retrieved wavefields. 

Once $F_{Ua}$ is resolved, we may evaluate the Marchenko equation again with the (conventional) workflow that was described in the previous section. However, this time we include $F_{Ua}$ in the left-hand side of (\ref{eq:S17}), such that we can retrieve (estimates of) waveforms in the coda $F_{Um}$ beyond the direct-wave approximation.

\section{Numerical example}
\label{numerical}

\begin{figure}
\centering
{\scalebox{0.80}[0.80]{\includegraphics[trim={0 0 0 0},clip]{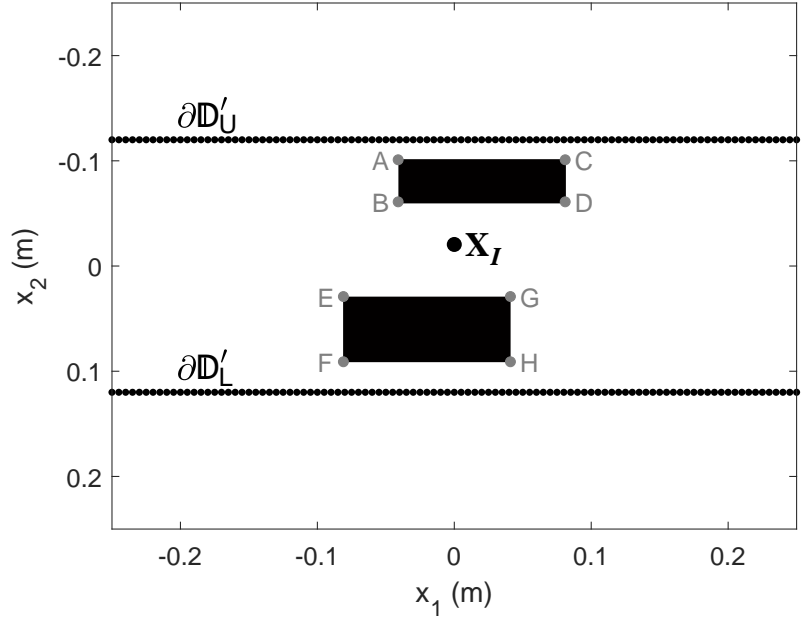}}}
\caption{Configuration for the 2D numerical experiment. The medium has a constant velocity of $1500 m/s$ and a background density $\rho_{white}=1000 kg/m^3$. The two black rectangles are density anomalies with $\rho_{black}=2000 kg/m^3$. The gray dots (labeled as $\mathrm{A}$, $\mathrm{B}$, ..., $\mathrm{H}$) denote sharp corners that generate forward-scattered waveforms. The upper boundary $\partial \mathbb{D}_U^\prime$ contains 101 coinciding vertical force (dipole) sources and pressure receivers with a spacing of $5mm$. The lower boundary $\partial \mathbb{D}_L^\prime$ contains 101 vertical force (dipole) sources with a spacing of $5mm$. Additional taper zones have been added to these arrays at the intervals $x_1 \in \left[-0.5m,-0.25m\right)$ and $x_1 \in \left(-0.25m,0.5m\right]$ to eliminate truncation artefacts of the spatial integrals. The black dot marks the specific location ${\bf{x}}_I=\left(0,-0.0205m \right)$, where we retrieve our focusing function and Green's function.
\label{fig:n3}}
\end{figure}

In this section, we apply the proposed methodology to ultrasonic data from a 2D numerical experiment in the $0-150kHZ$ frequency range. In Fig. \ref{fig:n3}, we show our synthetic model, which contains two rectangular density contrasts. These contrasts have been intentionally designed such that their corners (labeled as $\mathrm{A}, \mathrm{B}, ..., \mathrm{H}$) and vertical interfaces (labeled as $\mathrm{AB}$, $\mathrm{CD}$, $\mathrm{EF}$, $\mathrm{GH}$) generate forward-scattered waveforms that are not handled well by the conventional Marchenko methodology, which we want to improve on by incorporating auxiliary transmission data. Our aim is to retrieve $F_U$, $\Gamma_U$ and $\Gamma_L$ at the specific location ${\bf{x}}_I = \left( 0, -0.0205m\right)$, which is indicated in Fig. \ref{fig:n3}.

\begin{figure}
\centering
{\scalebox{0.80}[0.80]{\includegraphics[trim={0 0 0 0},clip]{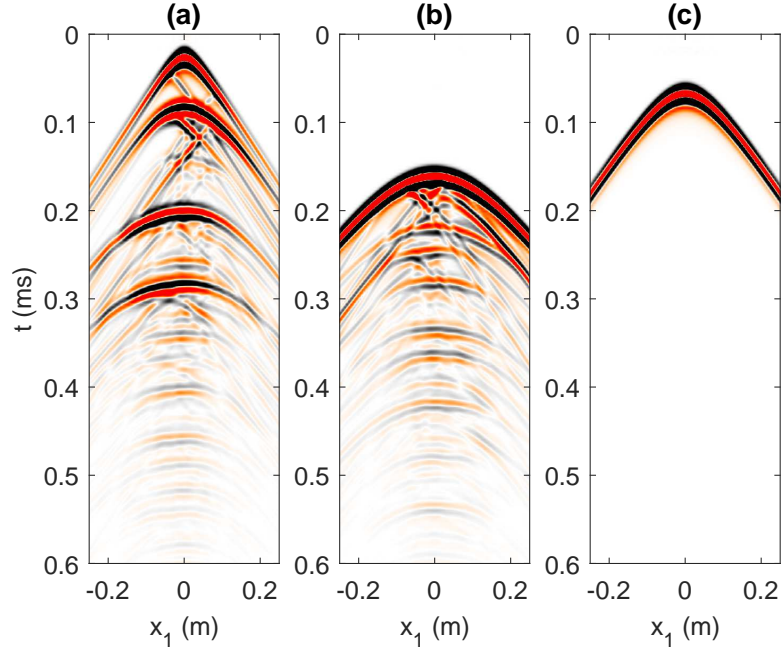}}}
\caption{
(a) Reflection response: $\forall {\bf{x}}_U^\prime \in \partial \mathbb{D}_U^\prime : \Gamma^- \left( {\bf{x}}_0, {\bf{x}}_U^\prime, t\right)$. (b) Transmission response: $\forall {\bf{x}}_L^\prime \in \partial \mathbb{D}_L^\prime : \Gamma^- \left( {\bf{x}}_0, {\bf{x}}_L^\prime, t\right)$. In both panels, we have set ${\bf{x}}_0=\left( 0, -0.120m \right)$ as a reference location. (c) Analytic dipole Green's function in a homogeneous background medium: $\forall {\bf{x}}_U^\prime \in \partial \mathbb{D}_U^\prime : \Gamma_0 \left( {\bf{x}}_I, {\bf{x}}_U^\prime, t\right)$.  All panels are convolved with the wavelet and clipped at 10\% of the maximum amplitude.
\label{fig:n4}}
\end{figure}

As a wavelet, we take the second derivative of a Gaussian function (also known as a Ricker wavelet) with a $50kHz$ peak frequency (for the truncation operators, we choose  $t_\epsilon = 20 \mu s$). Our traces consist of 1024  time samples with  $dt=3 \frac{1}{3}\mu s$. Reflection and transmission data are generated by solving an interface integral equation \cite{berg21}, for which we discretize the mass density model on a spatial grid with a spacing of $2mm$. In Fig. \ref{fig:n4}, we show sections of these data, as well as an analytic Green's function $\Gamma_{0}$ that we computed in a background homogeoneous medium.

\begin{figure}
\centering
{\scalebox{0.80}[0.80]{\includegraphics[trim={0 0 0 0},clip]{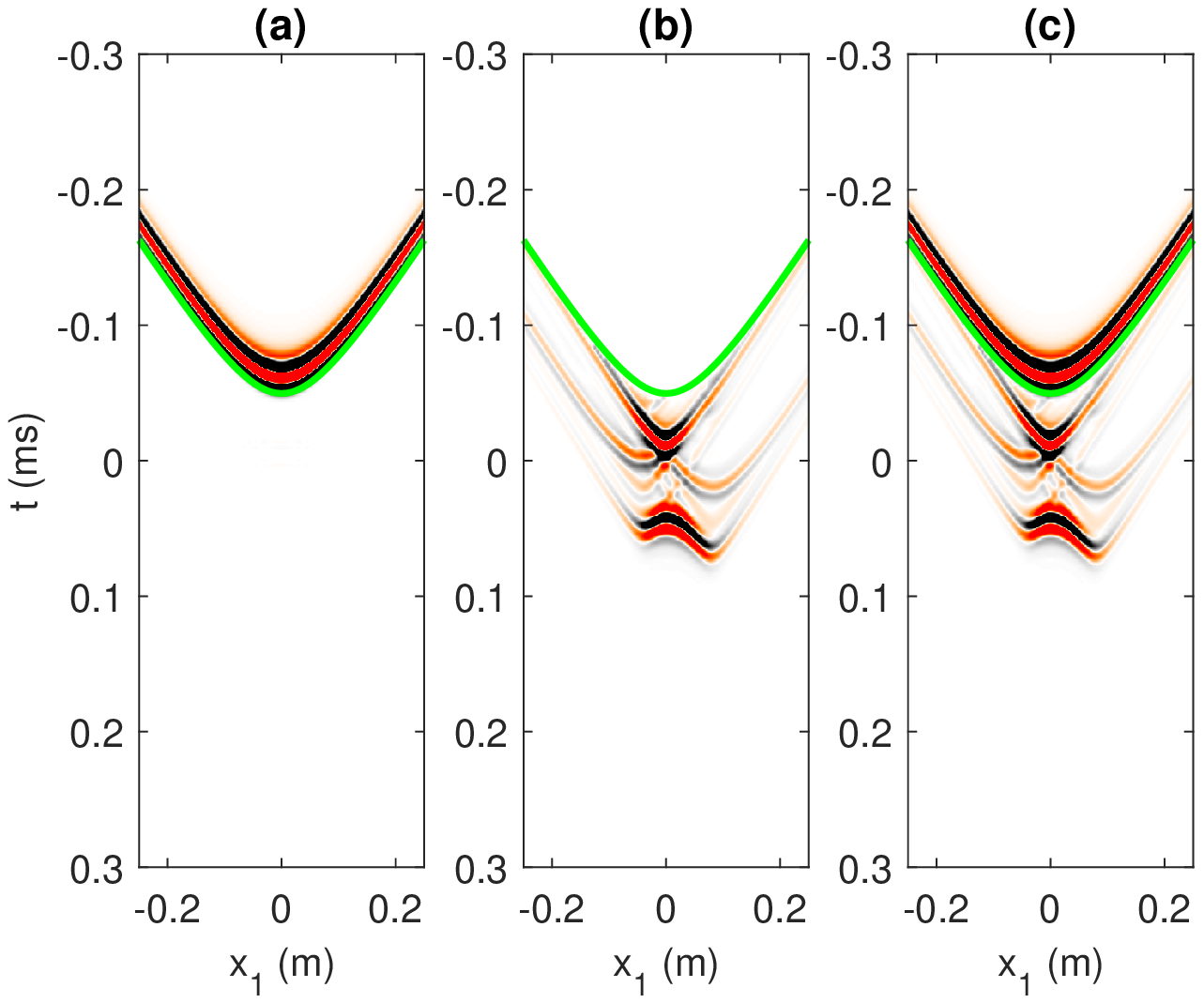}}}
\caption{
(a) Direct focusing function $F_{Ud}$. (b) Coda of the focusing function $F_{Um}$ as retrieved by solving the Marchenko equation under the direct-wave approximation $F_{Ua} = 0$. (c) Updated focusing function $F_{U}$, as obtained by adding the retrieved coda $F_{Um}$ to $F_{Ud}$. All panels are clipped at 10\% of the maximum amplitude of the direct wave.
\label{fig:n5}}
\end{figure}

For the estimation of the direct focusing function, we would like to reverse the analytic Green's function $\Gamma_0$ from \ref{fig:n4}(c) in time. Unfortunately, this procedure does not take care of transmission loss, which leads to an unacceptable amplitude error. Although methodologies exist to predict the transmission losses from the data \cite{mildner19b} (even in angle-dependent mode), we mitigate this problem here with help of a single (angle-independent) scaling factor $\alpha$, which is determined by

\begin{equation}
\alpha \left( {\bf{x}}_I \right)
=
\frac{
\int
\int_{\partial \mathbb{D}_U}
\Gamma_{0} \left( {\bf{x}}_I,{\bf{x}}_U, t \right)
G_0 \left( {\bf{x}}_U, {\bf{x}}_I, t \right)
{\rm d} {\bf{x}}_U
{\rm d} t
}
{
\int
\int_{\partial \mathbb{D}_U}
\Gamma_{0} \left( {\bf{x}}_I,{\bf{x}}_U, t \right)
G \left( {\bf{x}}_U, {\bf{x}}_I, t \right)
{\rm d} {\bf{x}}_U
{\rm d} t
}
.
\label{eq:S26}
\end{equation}
In the numerator of (\ref{eq:S26}), we focus a Green's function $G_0$ in the background medium at ${\bf{x}}_I$, while in the denominator, we do the same with a Green's function $G$ that is computed in the actual medium. The amplitude ratio between both focusing processes determines our scaling factor $\alpha$, which can be interpreted as an estimate of the average inverse transmission loss for propagation between $\partial \mathbb{D}_U^\prime$ and ${\bf{x}}_I$. By this procedure, we find $\alpha = 1.1141$ for the focal point that is specified in Fig. \ref{fig:n3}. The direct focusing function is then estimated as $F_{Ud} \approx \alpha \mathcal{Z} \Gamma_{0}$ and is shown in Fig. \ref{fig:n5}(a).

\begin{figure}
\centering
{\scalebox{1.00}[1.00]{\includegraphics[trim={0 0 0 0},clip]{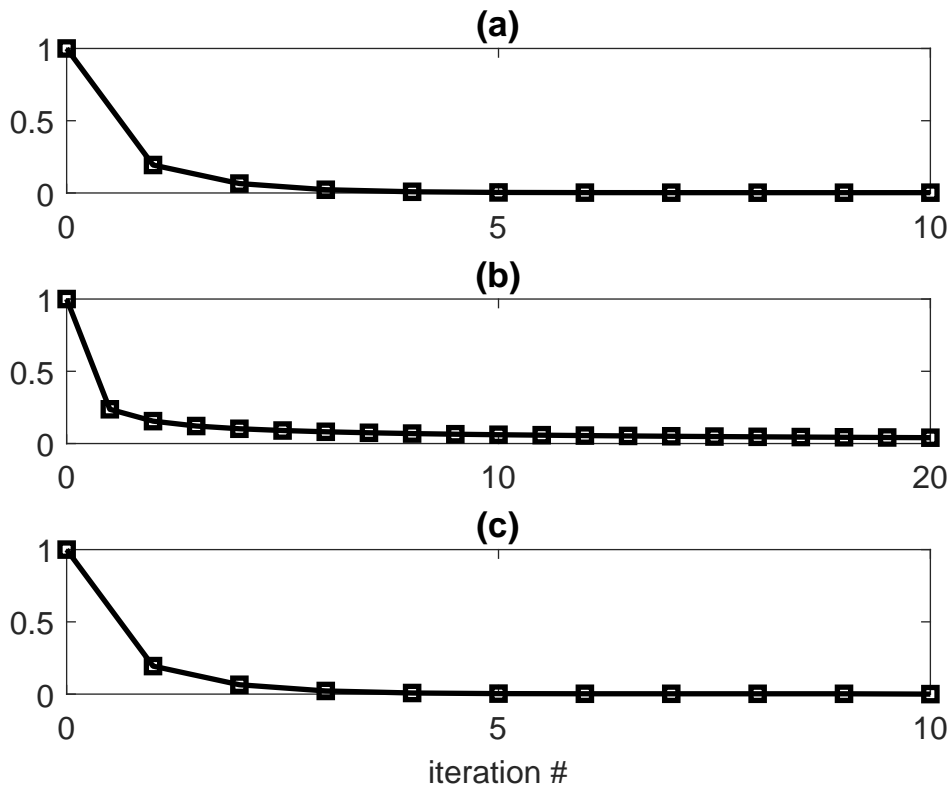}}}
\caption{
Frobenius norm of the residual (normalized by the residual at iteration $0$) when solving (a) the Marchenko equation under the direct-wave approximation $F_{Ua} =0$, (b) the auxiliary equation and (c) the Marchenko equation with an updated initial focusing function $F_{Ui}=F_{Ud}+F_{Ua}$.
\label{fig:n11}}
\end{figure}

First, we solve Marchenko equation (\ref{eq:S17}) in a conventional manner, using reflection data only. We do so by evaluating the first 10 terms of its associated Neumann series under the direct-wave approximation $F_{Ua} = 0$, see Fig. \ref{fig:n5}(b) (the convergence curve can be found in Fig. \ref{fig:n11}(a)). The retrieved coda $F_{Um}$ is added to $F_{Ud}$ and the result is shown in Fig. \ref{fig:n5}(c). 

\begin{figure}
\centering
{\scalebox{0.80}[0.80]{\includegraphics[trim={0 0 0 0},clip]{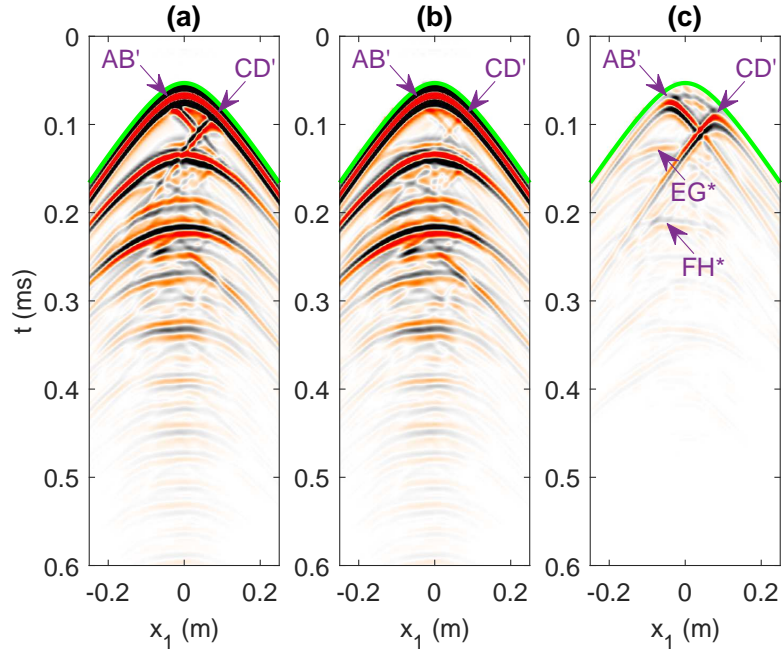}}}
\caption{
(a) Reference Green's function $\Gamma_U$, as obtained by forward modeling. (b) Same Green's function, as retrieved by solving Marchenko equation under the direct-wave approximation $F_{Ua} =0$. (c) Difference between Figs \ref{fig:n6}(b) and \ref{fig:n6}(a). The green curve marks the traveltime of  $t_{Ud}-t_\epsilon$. All panels are clipped at 10\% of the maximum amplitude of the direct wave.
\label{fig:n6}}
\end{figure}

For our reference, we compute the desired Green's function $\Gamma_U$ directly from the model parameters by solving an interface integral equation \cite{berg21}, see Fig. \ref{fig:n6}(a). In Fig. \ref{fig:n6}(b), we show  the same Green's function, as retrieved by the Marchenko equation under the direct-wave approximation $F_{Ua} =0$ (which is obtained by substituting the retrieved focusing function from Fig. \ref{fig:n5}(c) into (\ref{eq:S10})). The difference between the retrieved and reference Green's functions is given in Fig. \ref{fig:n6}(c). Events $\mathrm{AB}^\prime$ and $\mathrm{CD}^\prime$, which are indicated in Fig. \ref{fig:n6}, relate to forward-scattered waveforms that have not been retrieved accurately (where ${}^\prime$ is used to indicate forward-scattered waveforms). These events are mainly generated by the vertical interfaces $\mathrm{AB}$ and $\mathrm{CD}$ (as indicated in Fig. \ref{fig:n3}). A similar statement can be made about the (weaker) events $\mathrm{EG}^\star$ and $\mathrm{FH}^\star$, which seem to be related to the reflections of event $\mathrm{AB}^\prime$ at the horizontal interfaces $\mathrm{EG}$ and $\mathrm{FH}$ in Fig. \ref{fig:n3} (where ${}^\star$ is used to indicate reflected waveforms).

\begin{figure}
\centering
{\scalebox{0.80}[0.80]{\includegraphics[trim={0 0 0 0},clip]{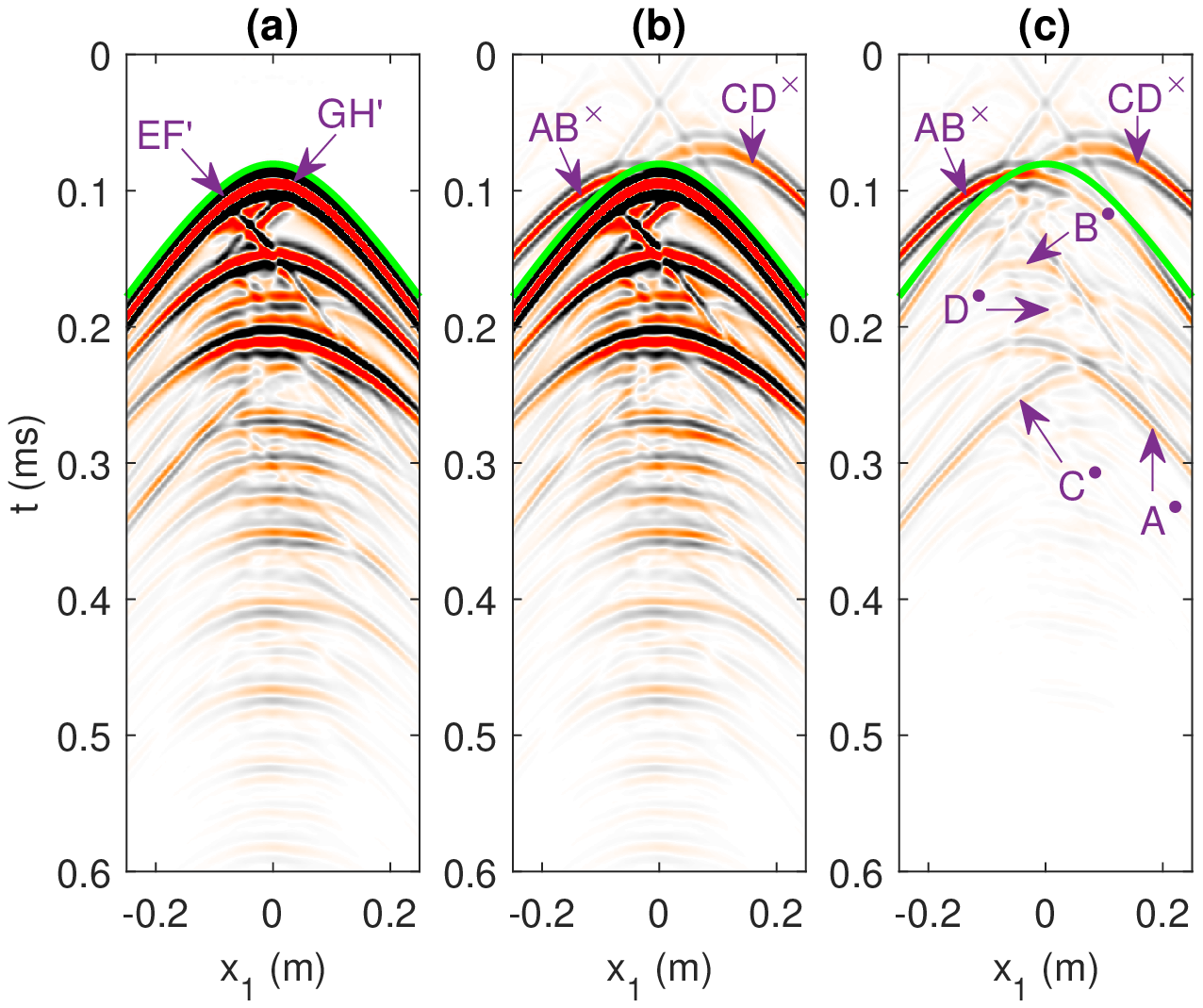}}}
\caption{
(a) Reference Green's function $\Gamma_L$, as obtained by forward modeling. (b) Same Green's function, as retrieved by solving Marchenko equation under the direct-wave approximation $F_{Ua} =0$. (c) Difference between Figs \ref{fig:n7}(b) and \ref{fig:n7}(a). The green curve marks the traveltime of $t_{Ld}-t_\epsilon$.  All panels are clipped at 10\% of the maximum amplitude of the direct wave.
\label{fig:n7}}
\end{figure}

The Green's function $\Gamma_L$ can be estimated by applying the transmission operator to the retrieved focusing function $F_U$, see (\ref{eq:S20}). In Fig. \ref{fig:n7}, we compare the result of this procedure with a reference Green's function that we found by solving an interface integral equation \cite{berg21}. Events $\mathrm{EF}^\prime$ and $\mathrm{GH}^\prime$ relate to forward-scattered waveforms that have been retrieved accurately. These events are mainly generated by the vertical interfaces $\mathrm{EF}$ and $\mathrm{GH}$ (as indicated in Fig. \ref{fig:n3}). Based on Fig. \ref{fig:n2}(a), it is clear that these waveforms map in the interval $\left[ t_{Ld} - t_\epsilon, \infty \right)$ (i.e. below the green curve in the figure). Hence, they will be in the nullspace of the operator $\Theta_L \mathcal{T}_{LU}$. On the other hand, events $\mathrm{AB}^\times$ and $\mathrm{CD}^\times$ are artefacts that are generated by forward-scattered waveforms from the vertical interfaces $\mathrm{AB}$ and $\mathrm{CD}$ in Fig. \ref{fig:n3} (where ${}^\times$ is used to indicate non-physical events). We confirm that these artefacts map mostly in the interval $\left( -\infty, t_{Ld} - t_\epsilon \right)$ (i.e. above the green curve in the figure), as we already predicted in Fig. \ref{fig:n2}(b). Hence, $\Theta_L \mathcal{T}_{LU} F_U \neq 0$, which violates (\ref{eq:S23}). This observation can be exploited to retrieve the missing components of the focusing function $F_{Ua}$ from the auxiliary equation, as we do shortly. Further, we have indicated $\mathrm{A}^\bullet$, $\mathrm{B}^\bullet$, $\mathrm{C}^\bullet$ and $\mathrm{D}^\bullet$ in Fig. \ref{fig:n7}(c). We interpret these as diffractions (indicated by ${}^\bullet$) from the corners $\mathrm{A}$, $\mathrm{B}$, $\mathrm{C}$ and $\mathrm{D}$ in Fig. \ref{fig:n3}, which have not been retrieved accurately.

\begin{figure}
\centering
{\scalebox{0.80}[0.80]{\includegraphics[trim={0 0 0 0},clip]{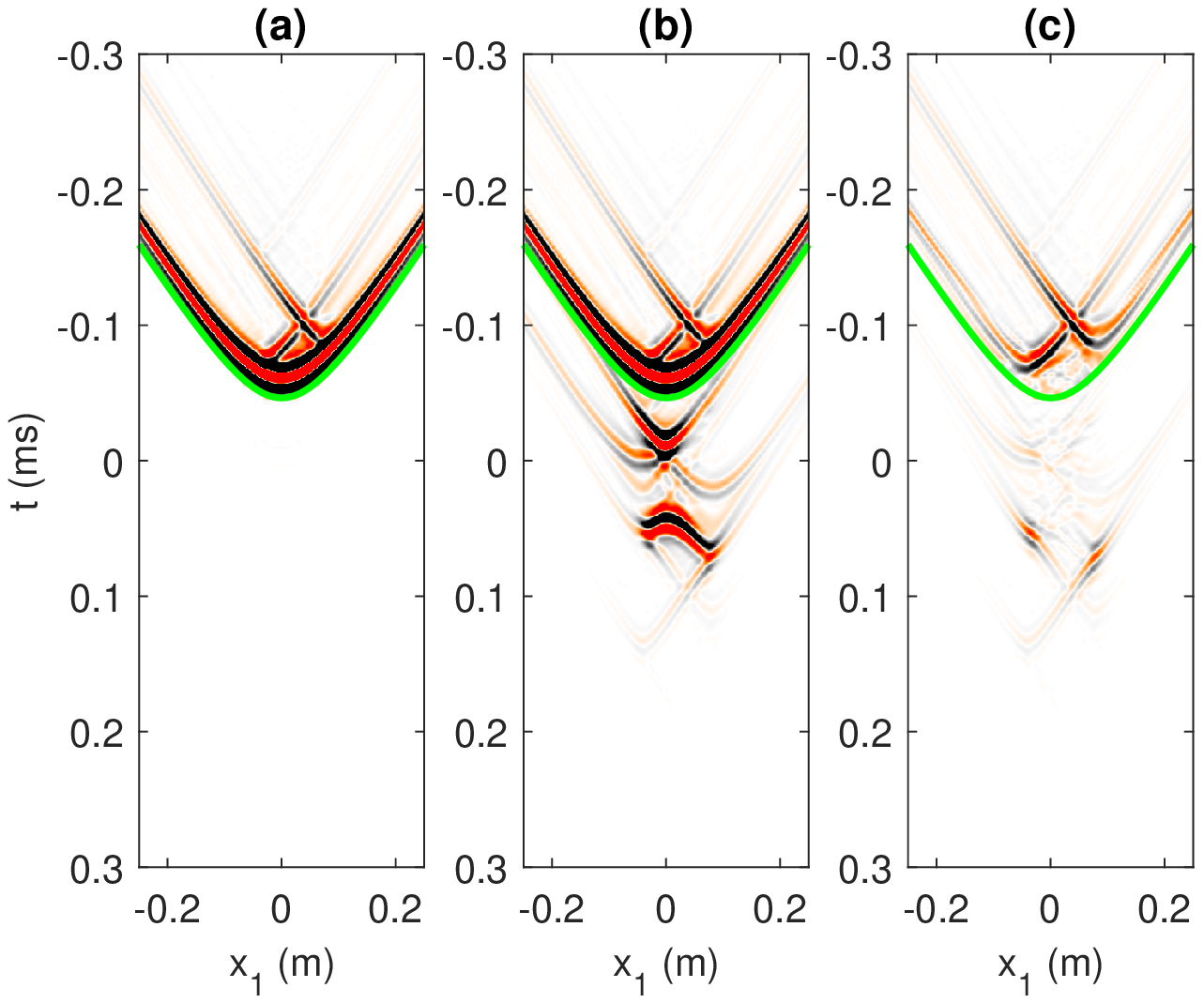}}}
\caption{
(a) Initial focusing function $F_{Ui}$ as obtained by adding $F_{Ua}$ (which is retrieved by least-squares inversion of (\ref{eq:S25})) to $F_{Ud}$. (b) Focusing function $F_U$ as retrieved by solving the Marchenko equation with $F_{Ui}=F_{Ud}+F_{Ua}$. (c) Difference between Figs \ref{fig:n8}(b) and \ref{fig:n5}(c).  Further, all settings are similar as in Fig. \ref{fig:n5}.
\label{fig:n8}}
\end{figure}

Next, we show how the missing components of the focusing function can be resolved from the transmission data. To achieve this, we solve (\ref{eq:S25}) for  $F_{Ua}$ on the interval $\left( -\infty, -t_{Ud} + t_\epsilon \right]$ by 20 iterations of LSQR \cite{paige82} (the convergence curve is provided in Fig. \ref{fig:n11}(b)). We add our solution to $F_{Ud}$ and show the result in Fig. \ref{fig:n8}(a). Then, we use $F_{Ud}$ and $F_{Ua}$ to invert Marchenko equation (\ref{eq:S17}) for $F_{Um}$  by evaluating the first 10 terms of its associated Neumann series (the convergence curve is provided in Fig. \ref{fig:n11}(c), which is not too different from  Fig. \ref{fig:n11}(a)). This results in a renewed estimate of the focusing function $F_U = F_{Ud} +F_{Ua} + F_{Um}$, see Fig. \ref{fig:n8}(b). In Fig. \ref{fig:n8}(c), we show the difference between Figs. \ref{fig:n8}(b) (obtained with transmissions) and \ref{fig:n5}(c) (obtained without transmissions). Several events can be observed in this figure, not only on the interval $\left( -\infty, -t_{Ud} + t_\epsilon \right]$ (i.e. updates of $F_{Ua}$), but also on $\left( -t_{Ud}+t_\epsilon, \infty \right)$ (i.e. updates of $F_{Um}$). These events help us to improve the retrieval of forward-scattered waveforms and their associated multiple reflections in the Green's functions $\Gamma_U$ and $\Gamma_L$, as we demonstrate next.

\begin{figure}
\centering
{\scalebox{0.80}[0.80]{\includegraphics[trim={0 0 0 0},clip]{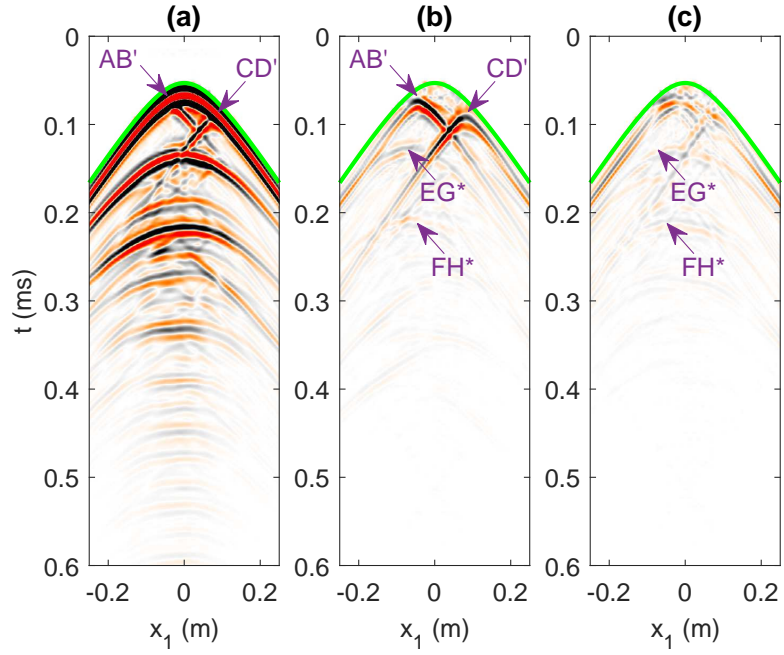}}}
\caption{
(a) Green's function $\Gamma_U$, as retrieved by solving the auxiliary equation and the Marchenko equation successively.
(b) Difference between Figs \ref{fig:n9}(a) and \ref{fig:n6}(b). (c) Difference between Figs \ref{fig:n9}(a) and \ref{fig:n6}(a). Further, all settings are similar as in Fig. \ref{fig:n6}.
\label{fig:n9}}
\end{figure}

In Fig. \ref{fig:n9}(a), we show the Green's function $\Gamma_U$, as obtained by substituting our renewed estimate of the focusing function into (\ref{eq:S10}). In Fig. \ref{fig:n9}(b),we show the difference between Figs. \ref{fig:n9}(a) (obtained with transmissions) and \ref{fig:n6}(b) (obtained without transmissions). As indicated in this figure, the forward-scattered waveforms that were indicated as $\mathrm{AB}^\prime$ and $\mathrm{CD}^\prime$, as well as $\mathrm{EG}^\star$ and $\mathrm{FH}^\star$ (relating to higher-order scattering), can be recognized. In Fig. \ref{fig:n9}(c), we show the difference between Fig. \ref{fig:n9}(a) and the reference Green's function in Fig. \ref{fig:n6}(a). By comparing this result with Fig. \ref{fig:n6}(c), we see that the forward-scattered waveforms ($\mathrm{AB}^\prime$ and $\mathrm{CD}^\prime$) and some of their associated multiples ($\mathrm{EG}^\star$ and $\mathrm{FH}^\star$) have been better resolved. We also observe that our result is not optimal, which we assume to be attributed to the incorrect amplitude spectrum of the direct focusing function (where transmission losses have not been handled well) and finite aperture.

\begin{figure}
\centering
{\scalebox{0.80}[0.80]{\includegraphics[trim={0 0 0 0},clip]{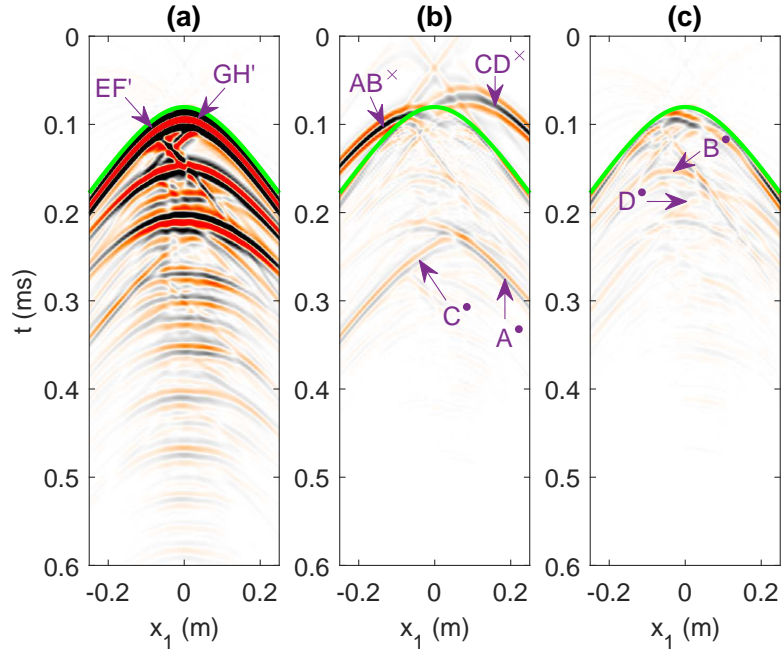}}}
\caption{(a) Green's function $\Gamma_L$, as retrieved by solving the auxiliary equation and the Marchenko equation successively. (b) Difference between Figs \ref{fig:n10}(a) and \ref{fig:n7}(b). (c) Difference between Figs \ref{fig:n10}(a) and \ref{fig:n7}(a). Further, all settings are similar as in Fig. \ref{fig:n7}.
\label{fig:n10}}
\end{figure}

We may also retrieve the Green's function $\Gamma_L$ by substituting our renewed estimate of the focusing function into (\ref{eq:S20}). The result of this operation (including the associated difference plots) is shown in Fig. \ref{fig:n10}. Comparing this result with Fig. \ref{fig:n7}, it is clear that the artefacts $\mathrm{AB}^\times$ and $\mathrm{CD}^\times$ have been suppressed, as enforced by the inversion of (\ref{eq:S25}). We also observe that the diffractions $\mathrm{A}^\bullet$ and $\mathrm{C}^\bullet$ have been retrieved better, which is not so much the case for the diffractions $\mathrm{B}^\bullet$ and $\mathrm{D}^\bullet$. The latter difference might be attributed to finite aperture. More specifically: the Fresnel zones \cite{wapenaar10} that are required for the retrieval of $\mathrm{B}^\bullet$ and $\mathrm{D}^\bullet$ are more extended and closer to the edges of the array compared to the those of $\mathrm{A}^\bullet$ and $\mathrm{C}^\bullet$. Consequently, they seem to suffer more from the finite aperture of the acquisition array.

\newpage
\section{Application to inverse source problems}

As a potential application of the proposed method, we focus our attention on inverse source problems, which are common in photoacoustic imaging \cite{huang13}. We consider the configuration in Fig. \ref{fig:n1} and assume that a distribution of sources is located inside the volume $\mathbb{D}$. We characterize this source distribution as $q_e \left( {\bf{x}},t \right)$, where subscript $e$ stands for ``experiment''. An important assumption is that all sources are ignited at $t=0$, with a zero-phase wavelet (which might have to be enforced in practice by additional preprocessing). Let  $\forall {\bf{x}}_U \in \partial \mathbb{D}_U : p_e \left( {\bf{x}}_U,t \right)$ be the recorded wavefield at the upper boundary. Our goal is to reconstruct the initial pressure distribution $\forall {\bf{x}} \in \mathbb{D} : p_e \left( {\bf{x}}, t=0 \right)$ from these recordings. A common strategy is to propagate the data back into the volume with Green's functions from a (typically smooth) macro model \cite{xu05}. However, a range of alternative solutions exists \cite{poudel19}. When reflection and transmission data are available, we may compute focusing functions by the methodology that we have derived in this paper and rely on (\ref{eq:S4}) for the required wavefield reconstruction process.  Unfortunately, (\ref{eq:S4}) can only be applied to a wavefield $p$ if its associated source distribution $q$ is zero throughout the volume $\mathbb{D}$, which is obviously not the case for $p = p_e$ and $q = q_e$.  To overcome this problem, we symmetrize the wavefield by the following operation 

\begin{equation}
p_h
\left( {\bf{x}},  t \right)
=
p_e
\left( {\bf{x}},  t \right)
+
p_e
\left( {\bf{x}},  -t \right)
.
\label{eq:S27}
\end{equation}
Here, subscript $h$ stands for ``homogeneous'', referring to the fact that $p_h$  is a solution to the homogeneous (= source-free) wave equation \cite{oristaglio89,wapenaar18}; i.e.  $p_h$  satisfies (\ref{eq:S2}) with $q =0$. Consequently, we can use (\ref{eq:S4}) to reconstruct $p_h$ throughout volume ${\mathbb{D}}$, given our recorded data at the upper boundary $\partial \mathbb{D}_U$. To facilitate this process, we substitute $p=p_h$ into (\ref{eq:S4}) and rewrite the result in the time domain as

\begin{equation}
\begin{split}
p_h \left( {\bf{x}} , t \right)
 =
\int_{\partial \mathbb{D}_U}
F_U \left( {\bf{x}},{\bf{x}}_U, t \right)
\ast
p_h^- \left( {\bf{x}}_U, t \right)
{\rm d} {\bf{x}}_U
+
\int_{\partial \mathbb{D}_U}
F_U \left( {\bf{x}},{\bf{x}}_U, -t \right)
\ast
p_h^+ \left( {\bf{x}}_U, t \right)
{\rm d} {\bf{x}}_U
.\end{split}
\label{eq:S28}
\end{equation}

Next, we realize that $\forall {\bf{x}}_U \in \partial \mathbb{D}_U : p_h^- \left( {\bf{x}}_U, t \right) = p_e \left( {\bf{x}}_U, t \right)$ and $p_h^+ \left( {\bf{x}}_U, t \right) = p_e \left( {\bf{x}}_U,- t \right)$. Substitution into (\ref{eq:S28}) yields

\begin{equation}
\begin{split}
p_h \left( {\bf{x}} , t \right)
 =
\int_{\partial \mathbb{D}_U}
F_U \left( {\bf{x}},{\bf{x}}_U, t \right)
\ast
p_e \left( {\bf{x}}_U, t \right)
{\rm d} {\bf{x}}_U
+
\int_{\partial \mathbb{D}_U}
F_U \left( {\bf{x}},{\bf{x}}_U, -t \right)
\ast
p_e \left( {\bf{x}}_U, -t \right)
{\rm d} {\bf{x}}_U
.\end{split}
\label{eq:S29}
\end{equation}
We can utilize this result to reconstruct the symmetrized data $p_h$ throughout the volume $\mathbb{D}$ from the recordings $p_e$ at the boundary $\partial \mathbb{D}_U$. To retrieve the initial pressure field, we may evaluate

\begin{equation}
p_e
\left( {\bf{x}},  t=0 \right)
=
\frac{1}{2}
p_h
\left( {\bf{x}},  t=0 \right)
,
\label{eq:S30}
\end{equation}
where we used (\ref{eq:S27}). In the following, we will demonstrate the potential of the proposed strategy with a simple synthetic example, based on the configuration that we presented earlier in Fig. \ref{fig:n3}. We choose the following (point) source distribution: $q_e \left( {\bf{x}},t \right) = \delta \left( {\bf{x}} - {\bf{x}}_I \right) S \left( t \right)$, where ${\bf{x}}_I= \left( 0,-0.0205m \right)$ and $S \left(t \right)$ is the second derivative of a Gaussian wavelet with a $50kHz$ peak frequency, as in our previous example. We compute the response $\forall {\bf{x}}_U \in {\partial \mathbb{D}}_U : p_e \left( {\bf{x}}_U, t \right)$ of this source distribution by forward modeling. Then, we reconstruct the symmetrized wavefield $\forall {\bf{x}} \in {\mathbb{D}} : p_h \left( {\bf{x}}, t \right)$ by  (\ref{eq:S29}), with help of focusing functions that we retrieve from reflection and transmission data. Finally, we evaluate  $\frac{1}{2} p_h \left( {\bf{x}},t=0 \right)$ to estimate the initial pressure field $p_e \left( {\bf{x}},t=0 \right)$, following (\ref{eq:S30}).  

\begin{figure}
\centering
{\scalebox{1.20}[1.20]{\includegraphics[trim={0 0 0 0},clip]{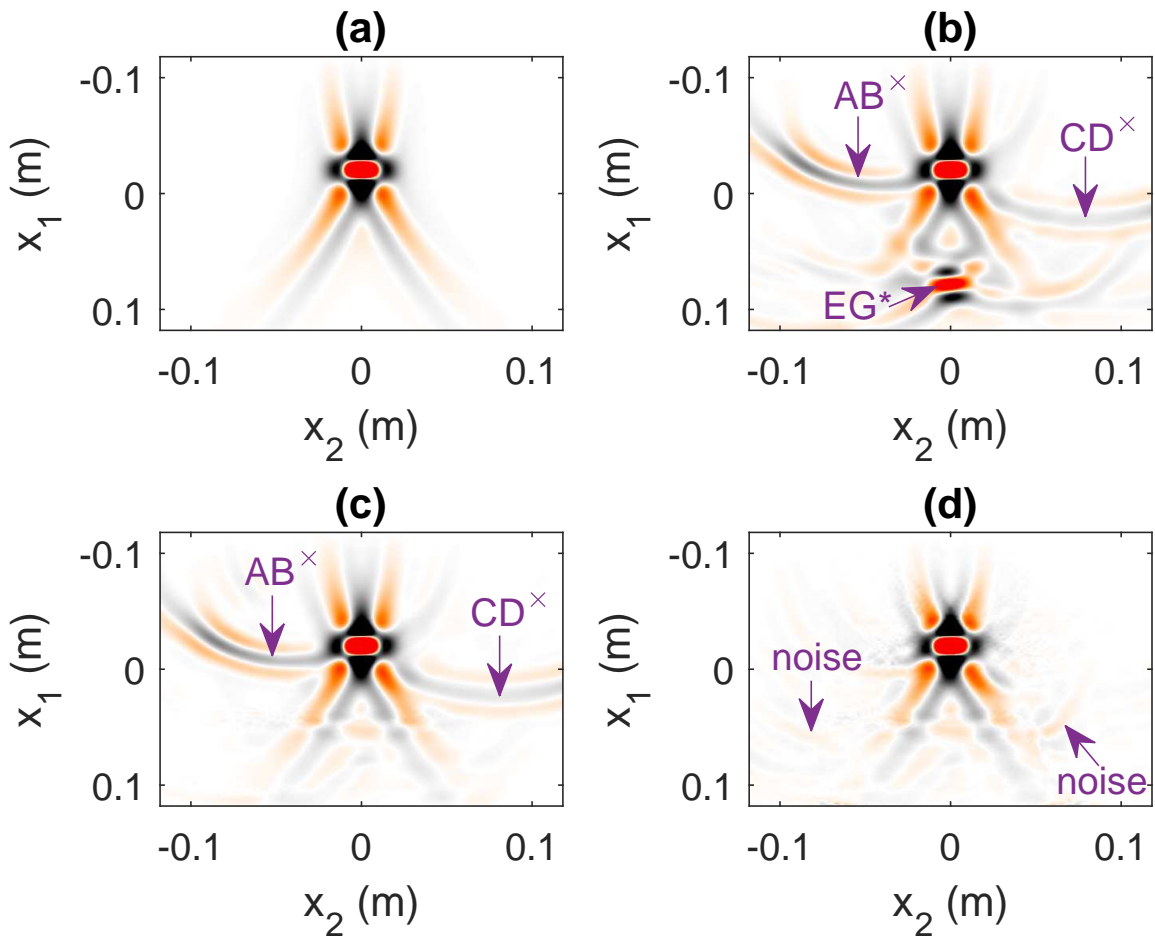}}}
\caption{Reconstructed initial pressure field $\forall {\bf{x}} \in \mathbb{D} : p_e \left( {\bf{x}},t=0 \right)$. In panel (a) (which serves as a reference for the other panels), data have been computed in a homogeneous background medium and $F_{Ud}$ acted as our focusing function. In panels (b)-(d), data have been computed in the heterogeneous medium of Fig. \ref{fig:n3}. (b) $F_{Ud}$ acted as our focusing function. (c) The focusing function is obtained by solving the Marchenko equation under the direct-wave approximation $F_{Ua} = 0$. (d) The focusing function is obtained by solving the auxiliary equation (with $F_{Um} = 0$) and the Marchenko equation successively. All panels are clipped at 10\% of the maximum amplitude.
\label{fig:n12}}
\end{figure}

For reference, we start with data $\forall {\bf{x}}_U \in \partial \mathbb{D}_U : p_e \left( {\bf{x}}_U, t \right)$ that was computed in a homogeneous medium with $c=1500 m/s$ and $\rho=1000 kg/m^3$ (i.e. without mass density contrast). We reconstruct the initial pressure distribution from these data, where we let  $F_{Ud}$ act as our focusing function. The result of this procedure is shown in Fig. \ref{fig:n12}(a). Since the model has not generated any scattered or reflected waveform, $F_{Ud}$ is identical to the complete focusing function (i.e. both $F_{Ua}$ and $F_{Um}$ are zero in this case). Consequently, our initial result cannot be improved by any of the approaches that we have discussed in this paper. Next, we repeat the exercise for data that were computed in the heterogeneous medium of Fig. \ref{fig:n3}, but we still choose $F_{Ud}$ as our focusing function for the evaluation of (\ref{eq:S29}). The image that is obtained by this procedure is shown in Fig. \ref{fig:n12}(b). Compared to Fig. \ref{fig:n12}(a), we observe two kinds of artefacts. Artefact $\mathrm{EG}^\star$ is related to reflections from the horizontal interface $\mathrm{EG}$ (see Fig. \ref{fig:n3}). This phenomenon can be understood as follows: the initial radiation from ${\bf{x}}_I$ has reflected at the interface $\mathrm{EG}$; this reflection was recorded at $\partial \mathbb{D}_U$ and has been backpropagated by $F_{Ud}$, generating a mirror image below ${\bf{x}}_I$. Artefacts $\mathrm{AB}^\times$ and $\mathrm{CD}^\times$ are related to forward scattering. These phenomena can be understood as follows: the initial radiation from ${\bf{x}}_I$ has interacted with the (corners and vertical interfaces of) the density contrast above ${\bf{x}}_I$, generating forward-scattered waveforms. These waveforms have been backpropagated by $F_{Ud}$ (where forward-scattered components have not been accounted for), resulting in additional events / artefacts (all intersecting at ${\bf{x}}_I$). Next, we solve the Marchenko equation under the direct-wave approximation $F_{Ua} = 0$ and we use the retrieved focusing functions to evaluate (\ref{eq:S29}). This leads to the image in Fig. \ref{fig:n12}(c). Note that the reflection-based artefact $\mathrm{EG}^\star$ has been removed by this procedure, but the artefacts from forward scattering remain. Finally, we compute the focusing functions by solving the auxiliary equation and the Marchenko equation successively, leading to the image in Fig. \ref{fig:n12}(d). This time, both types of artefacts have been suppressed significantly. It is also observed that some weaker artifacts have emerged in the image, which is indicated as noise in Fig. \ref{fig:n12}(d). To quantify the quality of our results, we introduce a global normalized root-mean-square error as

\begin{equation}
E
=
\sqrt{
\frac{\int_{\mathbb{D}}
\left|
I \left( {\bf{x}}\right)
-
I_0 \left( {\bf{x}}\right)
\right|^2 {\rm d} {\bf{x}}}
{\int_{\mathbb{D}}
\left|
I_0 \left( {\bf{x}}\right)
\right|^2 {\rm d} {\bf{x}}}
}.
\label{eq:S31}
\end{equation}
Here, $I$ is our image and $I_0$ is the associated reference image in Fig. \ref{fig:n12}(a). For the image $I$ in Fig. \ref{fig:n12}(b), we find $E=0.246$. This number is reduced significantly to $E=0.079$ for the image in Fig. \ref{fig:n12}(c), and down to $E=0.059$ in case of Fig. \ref{fig:n12}(d). This reduction of $E$ suggests that the result has improved. In our numerical example, we have chosen a monopole point source at ${\bf{x}}_I$  as our unknown source distribution. However, the linearity of the equations allows for arbitrary source distributions and mechanisms, as has been demonstrated in equivalent geophysical problems \cite{brackenhoff19b}. It is remarkable that - apart from the propagation velocity - no medium parameters are required for the reconstruction.

\section{discussion}

In this paper, we have proposed to partition the focusing function at an arbitrary location ${\bf{x}} \in {\mathbb{D}}$ as

\begin{equation}
F_U
=
F_{Ud}
+
F_{Ua}
+
F_{Um}.
\label{eq:S31}
\end{equation}
Here, $F_{Ud}$ is the direct focusing function, while $F_{Ua}$ and $F_{Um}$ contain all other waveforms  in the intervals $\left( -\infty, -t_{Ud} + t_\epsilon \right]$ and $\left( -t_{Ud} + t_\epsilon, \infty \right)$, respectively. We have assumed that $F_{Um}$ resides in the nullspace of $\Theta_L \mathcal{T}_{LU}$, such that it can be excluded from the auxiliary equation, while $F_{Ua}$ is not in this nullspace, and hence can be recovered. The latter assumption is motivated by Fig. \ref{fig:n2}(b), where we illustrate that forward-scattered waveforms which are generated above ${\bf{x}}$ are not in the nullspace of $\Theta_L \mathcal{T}_{LU}$. To investigate the validity of the abovementioned assumptions (especially in media with additional velocity contrast), it would be highly valuable to model $F_U$ directly from the medium parameters, for instance by depth extrapolation \cite{elison21}.

From the illustration in Fig. \ref{fig:n2}(b), it is clear that forward-scattered waveforms can only be retrieved from the auxiliary equation if they are sufficiently delayed (with respect to the direct wave). On a similar note, we have learned before that reflected waveforms can only be retrieved from the Marchenko equation if they are sufficiently delayed \cite{slob14}. It has been shown that the latter problem can be mitigated by introducing an augmented focusing function \cite{dukalski19,elison20,peng21a}. It might be worthwhile to investigate if a similar strategy could also be applied to the auxiliary equation.

In our study, we have not investigated the effects of finite spatial aperture, which can have a detrimental effect on Marchenko-based Green's function retrieval \cite{peng21,sripanich19b}. In our numerical examples, potential problems have been circumvented by choosing long spatial arrays and applying significant spatial tapers. With shorter arrays, it is well-understood that particular components of the Green's functions cannot be retrieved when the stationary points of the underlying integrals are not properly evaluated \cite{neut15}.

Further, we have chosen to demonstrate the validity of our workflow for 2D wave propagation only. Various publications have emerged recently on the implementation of the Marchenko equation for 3D wave propagation problems \cite{lomas20b,staring20,brackenhoff21}. Building on these developments, we prospect that a 3D implementation of the auxiliary equation should also be feasible.

We emphasize that our methodology has been derived under the assumption that the medium is lossless. By utilizing two-sided (rather than single-sided) reflection data and solving an alternative system of Marchenko equations, Green's functions can also be retrieved in dissipative media \cite{slob16}. A potential direction of further research is to include losses in our formulation of the auxiliary equation, such that forward-scattered waveforms and their associated multiples can also be retrieved in dissipative media. Such an approach might be beneficial not only for acoustic problems, but also for related applications that involve electromagnetic wave propagation, where losses typically cannot be neglected \cite{yang21}.

Last but not least, our work may have an impact on elastodynamic Green's function retrieval in solid media (in those fortunate cases where auxiliary transmission data are available). Although the underlying representations of the Marchenko equation are well established for elastodynamic wave propagation \cite{wapenaar14b,costa14}, it remains challenging to retrieve elastodynamic focusing functions in practice, since they overlap partly with their associated Green's functions in the time-space domain (due to the different velocities of P- and S-waves) and their forward-scattered components are generally unknown \cite{reinicke20}. It seems likely that these problems can be mitigated (at least to some extent) with help of auxiliary transmission data, by extending the theory from our paper to the elastodynamic case.

\section{Conclusion}

Marchenko-type focusing functions are useful tools to allow Green's function retrieval from single-sided reflection data. We can partition the focusing function in an intial focusing function and a coda. The initial focusing function contains the inverse direct wave and all preceding waveforms, which are typically associated with forward scattering. The coda contains all events after the inverse direct wave, which are typically associated with (primary and multiple) reflections. Given the initial focusing function, we can retrieve the coda from reflection data by solving a multidimensional Marchenko equation. In practice, the initial focusing function is often approximated by a time-reversed direct wave, which is generally computed in a macro velocity model. Under this (direct-wave) approximation, forward-scattered components of the Green's function and their associated multiples cannot be successfully recovered. In this paper, we have proposed to mitigate this problem by incorporating additional transmission data. We derived an auxiliary equation, which can be used to resolve the missing components of the initial focusing function from the transmission data by least-squares inversion. Once this is done, we can use our updated initial focusing function and the reflection data to solve the Marchenko equation beyond the direct-wave approximation. This procedure can be used to retrieve forward-scattered constituents of the Green's function in a medium with density contrast and constant propagation velocity, as we have demonstrated numerically. Implementing this methodology in a medium with velocity contrast is more challenging and requires additional research to be conducted.

\section{Acknowledgements}

This research has received funding from the European Research Council (grant no. 742703).

\bibliographystyle{IEEEtran}

\end{document}